\documentclass[preprint,showpacs,preprintnumbers,amsmath,amssymb]{revtex4}

\usepackage{graphicx}
\usepackage{epsfig}		
\usepackage{dcolumn}
\usepackage{bm}

\def\0{\mbox{\tiny $0$}}
\def\1{\mbox{\tiny $1$}}
\def\2{\mbox{\tiny $2$}}
\def\3{\mbox{\tiny $3$}}
\def\4{\mbox{\tiny $4$}}
\def\5{\mbox{\tiny $5$}}
\def\6{\mbox{\tiny $6$}}
\def\7{\mbox{\tiny $7$}}
\def\8{\mbox{\tiny $8$}}
\def\9{\mbox{\tiny $9$}}

\def\n{\mbox{\tiny $n$}}
\def\k{\mbox{\tiny $k$}}
\def\kk{\mbox{\small $k$}}

\def\f14{\mbox{\tiny $\frac{1}{4}$}}

\def\B{\mbox{\tiny $B$}}
\def\i{\mbox{\tiny $i$}}

\def\s{\mbox{\tiny $s$}}
\def\r{\mbox{\tiny $r$}}

\def\mi{\mbox{\tiny $-$}}
\def\ig{\mbox{\tiny $=$}}
\def\pl{\mbox{\tiny $+$}}
\def\ppm{\mbox{\tiny $\pm$}}

\def\bb#1{\mbox{\footnotesize $(#1)$}}

\begin{document}


\title{The construction of Dirac wave packets for a fermionic particle non-minimally coupling with an external magnetic field}

\author{A. E. Bernardini}
\affiliation{Department of Cosmic Rays and Chronology, State University of Campinas,\\
PO Box 6165, 13083-970, Campinas, SP, Brazil.}
\email{alexeb@ifi.unicamp.br}

\date{\today}

\begin{abstract}
We shall proceed with the construction of normalizable Dirac wave packets for {\em fermionic} particles (neutrinos) with dynamics governed by a ``modified'' Dirac equation with a non-minimal coupling with an external magnetic field.
We are not only interested on the analytic solutions of the ``modified'' Dirac wave equation but also on the construction of Dirac wave packets which can be used for describing the dynamics of some observable physical quantities which are relevant in the context of the quantum oscillation phenomena.
To conclude, we discuss qualitatively the applicability of this formal construction in the treatment of chiral (and flavor) oscillations in the theoretical context of neutrino physics.

\end{abstract}

\maketitle


Since those early days when Dirac had derived the relativistic wave equation for a free propagating
electron \cite{Dir28}, several efforts have been produced in the literature to solve the Dirac equation
with other analytical forms of interacting potentials, from central potential solutions \cite{Esp99,Alh05}
to recent theoretical attempts to describe quark confinement \cite{Ein81,Bak95}.
In fact, obtaining exact solutions of a generic class of Dirac wave equations becomes important since,
for many times, the conceptual understanding of physics can only be brought about by such solutions.
These solutions also correspond to valuable means for checking and improving models and numerical methods
for solving complicated physical problems.
In the context in which we intend to explore the Dirac formalism,
we can quote the effect of neutrino spin flipping attributed to some dynamic external \cite{Oli90}
interacting processes which come from the non-minimal coupling of a magnetic moment with an external electromagnetic
field \cite{Vol81} and which was formerly supposed to be a relevant effect in the context of the solar-neutrino puzzle
by suggesting an explanation for the LSND anomaly \cite{Ana98,Agu01,Ban03}.
To be more specific, our aim in this manuscript is to try to accommodate
the Dirac wave packet formalism \cite{Zub80,Ber04} and
a further class of static characteristics of neutrinos, namely, the (electro)magnetic
moment which appears in a Lagrangian with non-minimal coupling.
By following this line of reasoning, we are not only interested in solving a ``modified'' Dirac wave equation
but also in constructing Dirac wave packets which can be used for describing the dynamics of some observable
physical quantities which are relevant in the context of the quantum oscillation phenomena.

Despite their electric charge neutrality, neutrinos can interact with a photon through loop (radiative) diagrams \cite{Zub80,Kim93}.
The Lagrangian for the interaction between a fermionic field $\Psi\bb{x}$ and the electromagnetic
field $F^{\mu\nu}\bb{x}$ is given by
\begin{equation}
\mathcal{L} = \frac{1}{2}\,\mu\, \overline{\Psi}\bb{x}\, \sigma_{\mu\nu}F^{\mu\nu}\bb{x} \, \Psi\bb{x} + h.c.,
\end{equation}
with $x = \bb{t, \mbox{\boldmath$x$}}$, $\sigma_{\mu\nu} = \frac{i}{2}[\gamma_{\mu},\gamma_{\nu}]$ and $F^{\mu\nu}\bb{x}= \partial^{\mu}A^{\nu}\bb{x} - \partial^{\nu}A^{\mu}\bb{x}$ 
where we have not discriminated the flavor/mass mixing elements since we are initially interested just in
solving the equation of the motion described by 
\begin{eqnarray} 
i\frac{d}{dt}\Psi\bb{x} &=& \left\{\mbox{\boldmath$\alpha$}\cdot \mbox{\boldmath$p$} + \beta \left[m - \left(\frac{1}{2}\,\mu \,\sigma_{\mu\nu}F^{\mu\nu}\bb{x} + h.c.\right)\right]\right\}\Psi\bb{x}\nonumber\\
 		   &=& \left\{\mbox{\boldmath$\alpha$}\cdot \mbox{\boldmath$p$} + \beta \left[m - Re(\mu) \mbox{\boldmath$\Sigma$}\cdot \mbox{\boldmath$B$}\bb{x} +Im(\mu) \mbox{\boldmath$\alpha$}\cdot \mbox{\boldmath$E$}\bb{x}\right]\right\}\Psi\bb{x}
\label{08},~~
\end{eqnarray} 
where $\mbox{\boldmath$\alpha$} = \sum_{\k \ig \1}^{\3} \alpha_{\k}\hat{\kk} = \sum_{\k \ig \1}^{\3} \gamma_{\0}\gamma_{\k}\hat{\kk}$,
$\beta = \gamma_{\0}$, $\mbox{\boldmath$B$}\bb{x}$ and $\mbox{\boldmath$E$}\bb{x}$ are respectively the magnetic and electric fields.
The real (imaginary) part of $\mu$ represents the magnetic (electric) dipole moment of the mass eigenstate represented by
$\Psi\bb{x}$.
But it can be demonstrated \cite{Kim93} that, for Dirac neutrinos, the electric moment must vanish unless $CP$ is violated,
and, for Majorana neutrinos, the electric moment vanishes if $CPT$ invariance is assumed.
In the standard $SU(2)_{L} \otimes U(1)_{Y}$ electroweak theory \cite{Gla61}, if a positive chirality neutrino is an $SU(2)_{L}$ singlet,
the expression for $\mu$ can be found from Feynman diagrams for magnetic momentum corrections \cite{Kim93} and turns out to be
proportional to the neutrino mass (matrix),
\begin{equation} 
\mu = \frac{3\, e \,G}{8 \sqrt{2}\pi^{\2}} m = \frac{3\, m_e \,G}{4 \sqrt{2}\pi^{\2}}\, \mu_{\B} \, m_{\nu}
= 2.7 \times 10^{\mi \1\0}\,\mu_{\B}\,\frac{m_{\nu}}{m_N}
\end{equation} 
where $G$ is the Fermi constant and $m_{N}$ is the nucleon mass.
In principle, for $m_{\nu}\approx 1 \, eV$, the magnetic moment introduced by this formula is exceedingly small
to be detected or to affect astrophysical or physical processes.

From the theoretical point of view, the physical implications of the non-minimal coupling with an external magnetic field can then be studied by means of the
eigenvalue problem represented by the Hamiltonian equation
\begin{eqnarray} 
H\bb{\mbox{\boldmath$p$}} \, \Psi_{\n}\bb{\mbox{\boldmath$p$}} 
 		   &=& \left\{\mbox{\boldmath$\alpha$}\cdot \mbox{\boldmath$p$} + \beta \left[m - Re(\mu) \mbox{\boldmath$\Sigma$}\cdot \mbox{\boldmath$B$}\right]\right\}\Psi_{\n}\bb{\mbox{\boldmath$p$}}
\label{10},~~
\end{eqnarray} 
which represents the time evolution operator of a spinor $\Psi\bb{x}$
for times subsequent to the creation $(t = 0)$.
The most general solution to the above eigenvalue problem is represented by
\begin{eqnarray} 
E_{\n}\bb{\mbox{\boldmath$p$}} = \,  \pm E_{\s}\bb{\mbox{\boldmath$p$}} 
&=& \pm \sqrt{m^{\2} + \mbox{\boldmath$p$}^{\2} + \mbox{\boldmath$a$}^{\2} +\bb{\mi 1}^{\s}2
\sqrt{m^{\2}\mbox{\boldmath$a$}^{\2} + \bb{\mbox{\boldmath$p$} \times \mbox{\boldmath$a$}}^{\2}}}, ~~~~s\,=\, 1,\,2
\label{11},
\end{eqnarray} 
where $\mbox{\boldmath$a$} = Re(\mu) \mbox{\boldmath$B$}$ and the complete set of orthonormal eigenstates corresponds to
\begin{eqnarray} 
(\Psi^{\pl}\bb{p_{\s}})^{\dagger} &=& -N\bb{p_{\s}}\, \left\{\sqrt{\frac{A^{\mi}_{\s}}{A^{\pl}_{\s}}},\,\sqrt{\frac{\alpha^{\pl}_{\s}}{\alpha^{\mi}_{\s}}},\,\sqrt{\frac{A^{\mi}_{\s}\alpha^{\pl}_{\s}}{A^{\pl}_{\s}\alpha^{\mi}_{\s}}},\,-1\right\}\nonumber\\
(\Psi^{\mi}\bb{p_{\s}})^{\dagger} &=& -N\bb{p_{\s}}\, \left\{\sqrt{\frac{A^{\mi}_{\s}}{A^{\pl}_{\s}}},\,-\sqrt{\frac{\alpha^{\mi}_{\s}}{\alpha^{\pl}_{\s}}},\,-\sqrt{\frac{A^{\mi}_{\s}\alpha^{\mi}_{\s}}{A^{\pl}_{\s}\alpha^{\pl}_{\s}}},\,-1\right\}
\label{12},
\end{eqnarray} 
where $p_{\s}$ is the relativistic {\em quadrimomentum}, $p_{\s} = (E_{\s}\bb{\mbox{\boldmath$p$}}, \mbox{\boldmath$p$})$,
$N\bb{p_{\s}}$ is the normalization constant and
\begin{equation} 
A^{\ppm}_{\s} = \Delta_{\s}^{\2}\bb{\mbox{\boldmath$p$}} \pm 2 m |\mbox{\boldmath$a$}|  - \mbox{\boldmath$a$}^{\2},~~~
\alpha^{\ppm}_{\s}= 2 E_{\s}\bb{\mbox{\boldmath$p$}} |\mbox{\boldmath$a$}| \pm (\Delta_{\s}^{\2}\bb{\mbox{\boldmath$p$}} + \mbox{\boldmath$a$}^{\2})
\nonumber
\end{equation}
with
\begin{equation}
\Delta_{\s}^{\2}\bb{\mbox{\boldmath$p$}} = E{\s}^{\2}\bb{\mbox{\boldmath$p$}} - (m^{\2} + \mbox{\boldmath$p$}^{\2}) + \mbox{\boldmath$a$}^{\2}.
\end{equation}
We can observe that the components of $\Psi^{\ppm}\bb{p_{\s}}$ does not satisfy any auxiliary conditions,
namely, at a given time $t$, they are independent functions of $\mbox{\boldmath$p$}$ and the eigenspinors does not represent
neither chirality nor helicity eigenstates.
Thus, if we seek the plane wave decomposition as
\begin{eqnarray}
&&\Psi^{^{\pl}}\bb{x} = \exp{[- i(E_{\s}\bb{\mbox{\boldmath$p$}}\,t -\mbox{\boldmath$p$} \cdot \mbox{\boldmath$x$})]}
\,\mathcal{U}\bb{p_{\s}},
 ~~~~\mbox{for positive frequencies and}\nonumber\\
&&\Psi^{^{\mi}}\bb{x} = \exp{[~ i(E_{\s}\bb{\mbox{\boldmath$p$}}\,t -\mbox{\boldmath$p$} \cdot \mbox{\boldmath$x$})]}\,\mathcal{V}\bb{p_{\s}},
 ~~~~\mbox{for negative frequencies}
\label{13}
\end{eqnarray}
the time evolution of a plane wave packet solution $\Psi\bb{t, \mbox{\boldmath$x$}}$ in this circumstances can be written as
\begin{eqnarray}
\Psi\bb{t, \mbox{\boldmath$x$}}
&=& \int\hspace{-0.1 cm} \frac{d^{\3}\hspace{-0.1cm}\mbox{\boldmath$p$}}{(2\pi)^{\3}}
\sum_{\s \ig \1,\2}\{b\bb{p_{\s}}\mathcal{U}\bb{p_{\s}}\, \exp{[- i\,E_{\s}\bb{\mbox{\boldmath$p$}}\,t]}\nonumber\\
&&~~~~~~~~~~~~~~~~+ d^*\bb{\tilde{p}_{\s}}\mathcal{V}\bb{\tilde{p}_{\s}}\, \exp{[+i\,E_{\s}\bb{\mbox{\boldmath$p$}}\,t]}\}
\exp{[i \, \mbox{\boldmath$p$} \cdot \mbox{\boldmath$x$}]},~~~\mbox{with}~~ \tilde{p}_{\s} = (E_{\s},-\mbox{\boldmath$p$}),
\label{14}
\end{eqnarray}
which, however, requires some extensive mathematical manipulations
for explicitly constructing the dynamics of an operator $\mathcal{O}$ as
\begin{equation}
\mathcal{O}\bb{t} = \int{d^{\3}\mbox{\boldmath$x$}\,\Psi^{\dagger}\bb{t, \mbox{\boldmath$x$}}\,\mathcal{O}\,
\Psi\bb{t, \mbox{\boldmath$x$}}}
\label{15}.
\end{equation}
Meanwhile, if the quoted observables like the chirality $\gamma^{\5}$,
the helicity $\mbox{\boldmath$\Sigma$}\cdot\mbox{\boldmath$p$}$ or even the spin projection onto the magnetic field $\mbox{\boldmath$\Sigma$}\cdot\mbox{\boldmath$B$}$)
commute with the Hamiltonian $H$, we could reconfigure the above solutions to simpler ones. 
To illustrate this point we shall limit our analysis to very restrictive spatial configurations of $\mbox{\boldmath$B$}$ so that,
as a first attempt, we could calculate the observable expectation values which appear in Eq.(\ref{15}).
Let us then assume that the magnetic field $\mbox{\boldmath$B$}$ is either orthogonal or parallel to the momentum $\mbox{\boldmath$p$}$.
For both of these cases the spinor eigenstates can now be decomposed into bi-spinors which are eigenstates of the spin projection operator
$\mbox{\boldmath$\sigma$}\cdot\mbox{\boldmath$B$}$ as
\begin{equation}
\mathcal{U}\bb{p_{\s}} = N^{\pl}\bb{p_{\s}}\left[\begin{array}{r} \varphi^{\pl}\bb{p_{\s}}\\ \chi^{\pl}\bb{p_{\s}}\end{array}\right]
\end{equation}
and
\begin{equation} 
\mathcal{V}\bb{p_{\s}} = N^{\mi}\bb{p_{\s}}\left[\begin{array}{r} \varphi^{\mi}\bb{p_{\s}}\\ \chi^{\mi}\bb{p_{\s}}\end{array}\right]
\label{16},
\end{equation}
i. e. beside of being energy eigenstates, the general solutions $\mathcal{U}\bb{p_{\s}}$
and $\mathcal{V}\bb{p_{\s}}$ will become eigenstates of the operator $\mbox{\boldmath$\Sigma$}\cdot\mbox{\boldmath$B$}$ and, equivalently, of $\mbox{\boldmath$\Sigma$}\cdot\mbox{\boldmath$a$}$.
The Eq.(\ref{10}) can thus be decomposed into a pair of coupled equations like
\begin{eqnarray}
\left(\pm E_{\s} - m + \mbox{\boldmath$\sigma$}\cdot\mbox{\boldmath$a$} \right)\varphi^{\ppm}_{\s} &=& \pm \mbox{\boldmath$\sigma$}\cdot\mbox{\boldmath$p$}\,\chi^{\ppm}_{\s},\nonumber\\
\left(\pm E_{\s} + m - \mbox{\boldmath$\sigma$}\cdot\mbox{\boldmath$a$} \right)\chi^{\ppm}_{\s} &=& \pm \mbox{\boldmath$\sigma$}\cdot\mbox{\boldmath$p$}\,\varphi^{\ppm}_{\s},
\label{17}
\end{eqnarray}
where we have suppressed the $p_{\s}$ dependence.
The constraint imposed by the above equations leads to following representation for the energy eigenstates,
\begin{equation}
\mathcal{U}\bb{p_{\s}} = N^{\pl}\bb{p_{\s}}\left[\begin{array}{r} \varphi^{\pl}_{\s}\\ \frac{\left(E_{\s} + m + \mbox{\boldmath$\sigma$}\cdot\mbox{\boldmath$a$}\right)\mbox{\boldmath$\sigma$}\cdot\mbox{\boldmath$p$}}{\left[(E_{\s}+m)^{\2}+\mbox{\boldmath$a$}^{\2}\right]}\,\varphi^{\pl}_{\s}\end{array}\right]\end{equation} and \begin{equation} 
\mathcal{V}\bb{p_{\s}} = N^{\mi}\bb{p_{\s}}\left[\begin{array}{r} \frac{\left(E_{\s} + m + \mbox{\boldmath$\sigma$}\cdot\mbox{\boldmath$a$}\right)\mbox{\boldmath$\sigma$}\cdot\mbox{\boldmath$p$}}{\left[(E_{\s}+m)^{\2}+\mbox{\boldmath$a$}^{\2}\right]}\,\chi^{\mi}_{\s}\\ \chi^{\mi}_{\s}\end{array}\right]
\label{18}.
\end{equation}
Since  $\varphi^{\pl}_{\1,\2} \equiv \chi^{\mi}_{\1,\2} =\left(\begin{array}{r} 1 \\ 0\end{array}\right),\,
\left(\begin{array}{r} 0 \\ 1\end{array}\right)$
represent orthonormal eigenvectors of $\mbox{\boldmath$\sigma$}\cdot\mbox{\boldmath$a$}$, we can immediately deduce the orthogonality relations
\begin{eqnarray}
&&\mathcal{U}^{\dagger}\bb{p_{\s}} \, \mathcal{U}\bb{p_{\r}} = 
\mathcal{V}^{\dagger}\bb{p_{\s}} \, \mathcal{V}\bb{p_{\r}} = \delta_{\s\r},
\nonumber\\
&&\mathcal{U}^{\dagger}\bb{p_{\s}} \,\gamma_{\0}\, \mathcal{V}\bb{p_{\r}} = 
\mathcal{V}^{\dagger}\bb{p_{\s}} \,\gamma_{\0}\, \mathcal{U}\bb{p_{\r}} = 0
\label{20}.
\end{eqnarray}
The above results can be obtained by simply introducing the commuting relation 
$[\mbox{\boldmath$\sigma$}\cdot\mbox{\boldmath$p$},\,\mbox{\boldmath$\sigma$}\cdot\mbox{\boldmath$B$}] = 0$ 
which is derived when $\mbox{\boldmath$p$}\times\mbox{\boldmath$B$} = 0$ and the anti-commuting relation
$\{\mbox{\boldmath$\sigma$}\cdot\mbox{\boldmath$p$},\,\mbox{\boldmath$\sigma$}\cdot\mbox{\boldmath$B$}\}$ when 
$\mbox{\boldmath$p$}\cdot\mbox{\boldmath$B$} = 0$.
Now let us separately summarize other relevant information for both of these restrictive cases.
If $\mbox{\boldmath$p$}\cdot\mbox{\boldmath$B$} = 0$, the eigenspinor representation can be reduced to
\begin{equation}
\mathcal{U}\bb{p_{\s}} = \sqrt{\frac{\epsilon_{\0} + m}{2\epsilon_{\0}}}
\left[\begin{array}{r} \varphi^{\pl}_{\s}\\ \frac{\mbox{\boldmath$\sigma$}\cdot\mbox{\boldmath$p$}}{\epsilon_{\0}+ m}\,\varphi^{\pl}_{\s}\end{array}\right]\end{equation} and \begin{equation} 
\mathcal{V}\bb{p_{\s}} = \sqrt{\frac{\epsilon_{\0} + m}{2\epsilon_{\0}}}
\left[\begin{array}{r} \frac{\mbox{\boldmath$\sigma$}\cdot\mbox{\boldmath$p$}}{\epsilon_{\0}+ m}\,\chi^{\mi}_{\s}\\ \chi^{\mi}_{\s}\end{array}\right]
\label{21},
\end{equation}
with $\epsilon_{\0} = \sqrt{\mbox{\boldmath$p$}^{\2} + m^{\2}}$
and the energy eigenvalues
\begin{equation}
\pm E_{\s} = \pm\left[\epsilon_{\0} + \bb{\mi 1}^{\s}|\mbox{\boldmath$a$}|\right]
\label{22}.
\end{equation}
The closure relations can be constructed in terms of 
\begin{eqnarray}
\sum_{\s\,=\,\1,\2}{\mathcal{U}\bb{p_{\s}}\otimes\mathcal{U}^{\dagger}\bb{p_{\s}}\gamma_{\0}}&=&
\frac{\gamma_{\mu}p_{\0}^{\mu} + m}{2 \epsilon_{\0}} \sum_{\s\,=\,\1,\2}
{\left[\frac{1-\bb{\mi 1}^{\s}\gamma_{\0}\mbox{\boldmath$\Sigma$}\cdot\hat{\mbox{\boldmath$a$}}}{2}\right]}\nonumber\\
-\sum_{\s\,=\,\1,\2}{\mathcal{V}\bb{p_{\s}}\otimes\mathcal{V}^{\dagger}\bb{p_{\s}}\gamma_{\0}}&=&
\frac{-\gamma_{\mu}p_{\0}^{\mu} + m}{2 \epsilon_{\0}} \sum_{\s\,=\,\1,\2}
{\left[\frac{1-\bb{\mi 1}^{\s}\gamma_{\0}\mbox{\boldmath$\Sigma$}\cdot\hat{\mbox{\boldmath$a$}}}{2}\right]},
\label{23}
\end{eqnarray}
where $p_{\0} = (\epsilon_{\0}, \mbox{\boldmath$p$})$.
If $\mbox{\boldmath$p$} \times \mbox{\boldmath$B$} = 0$, the eigenspinor representation can be reduced to
\begin{equation}
\mathcal{U}\bb{p_{\s}} = \sqrt{\frac{E_{\s} + m_{\s}}{2E_{\s}}}
\left[\begin{array}{r} \varphi^{\pl}_{\s}\\ \frac{\mbox{\boldmath$\sigma$}\cdot\mbox{\boldmath$p$}}{E_{\s}+ m_{\s}}\,\varphi^{\pl}_{\s}\end{array}\right]\end{equation} and \begin{equation} 
\mathcal{V}\bb{p_{\s}} = \sqrt{\frac{E_{\s} + m_{\s}}{2E_{\s}}}
\left[\begin{array}{r} \frac{\mbox{\boldmath$\sigma$}\cdot\mbox{\boldmath$p$}}{E{\s}+ m_{\s}}\,\chi^{\mi}_{\s}\\ \chi^{\mi}_{\s}\end{array}\right]
\label{25},
\end{equation}
with $m_{\s} = m + \bb{\mi 1}^{\s}|\mbox{\boldmath$a$}|$
and the energy eigenvalues
\begin{equation}
\pm E_{\s} = \pm \sqrt{\mbox{\boldmath$p$}^{\2} + m_{\s}^{\2}}
\label{26}.
\end{equation}
In this case, the closure relations can be constructed in terms of 
\begin{eqnarray}
\sum_{\s\,=\,\1,\2}{\mathcal{U}\bb{p_{\s}}\otimes\mathcal{U}^{\dagger}\bb{p_{\s}}\gamma_{\0}}&=&
\sum_{\s\,=\,\1,\2}{\left\{\frac{\gamma_{\mu}p_{\s}^{\mu} + m_{\s}}{2 E_{\s}} 
\left[\frac{1-\bb{\mi 1}^{\s}\mbox{\boldmath$\Sigma$}\cdot\hat{\mbox{\boldmath$a$}}}{2}\right]\right\}}\nonumber\\
-\sum_{\s\,=\,\1,\2}{\mathcal{V}\bb{p_{\s}}\otimes\mathcal{V}^{\dagger}\bb{p_{\s}}\gamma_{\0}}&=&
 \sum_{\s\,=\,\1,\2}{\left\{\frac{-\gamma_{\mu}p_{\s}^{\mu} + m_{\s}}{2 E_{\s}} \left[\frac{1-\bb{\mi 1}^{\s}\mbox{\boldmath$\Sigma$}\cdot\hat{\mbox{\boldmath$a$}}}{2}\right]\right\}}.
\label{27}
\end{eqnarray}

Finally, the calculation of the expectation value of $\mathcal{O}\bb{t}$ is substantially simplified when
we substitute the above closure
relations into the wave packet expression of Eq.~(\ref{14}).
To clarify this point, we suppose
the initial condition over $\Psi\bb{0,\mbox{\boldmath$x$}}$
is the Fourier transform of the weight function 
\begin{equation}
\varphi\bb{\mbox{\boldmath$p$}-\mbox{\boldmath$p$}_{\i}}\,w \,=\,
\sum_{\s \ig \1,\2}{\{b\bb{p_{\s}}\mathcal{U}\bb{p_{\s}} + d^*\bb{\tilde{p}{\s}}\mathcal{V}\bb{\tilde{p}_{\s}}\}}
\label{28}
\end{equation}
so that
\begin{equation}
\Psi\bb{0, \mbox{\boldmath$x$}}
= \int\hspace{-0.1 cm} \frac{d^{\3}\hspace{-0.1cm}\mbox{\boldmath$p$}}{(2\pi)^{\3}}
\varphi\bb{\mbox{\boldmath$p$}-\mbox{\boldmath$p$}_{\i}}\exp{[i \, \mbox{\boldmath$p$} \cdot \mbox{\boldmath$x$}]}
\,w
\label{29}
\end{equation}
where $w$ is some fixed normalized spinor.
By using the orthogonality properties of Eq.~(\ref{20}),
we find \cite{Zub80}
\begin{eqnarray}
b\bb{p_{\s}} &=& \varphi\bb{\mbox{\boldmath$p$}- \mbox{\boldmath$p$}_{\i}} \, \mathcal{U}^{\dagger}\bb{p_{\s}} \, w, \nonumber\\
d^*\bb{\tilde{p}_{\s}} &=& \varphi\bb{\mbox{\boldmath$p$}- \mbox{\boldmath$p$}_{\i}}\,\mathcal{V}^{\dagger}\bb{\tilde{p}_{\s}}\, w.
\label{30}
\end{eqnarray}
These coefficients carry important physical information.
For {\em any} initial state $\Psi\bb{0, \mbox{\boldmath$x$}}$ given by Eq.~(\ref{29}),
the negative frequency solution coefficient
$d^*\bb{\tilde{p}_{\s}}$ necessarily provides a non-null contribution to the time evolving wave packet.
This obliges us to take the complete set of Dirac equation solutions to construct the wave packet.
Only if we consider the initial spinor $w$ being a positive energy ($E_{\s}\bb{\mbox{\boldmath$p$}}$) and momentum 
($\mbox{\boldmath$p$}$) eigenstate, the contribution due to the negative frequency solutions 
$d^*\bb{\tilde{p}_{\s}}$ will become null and we will have a simple expression for the time evolution of
any physical observable.

As we have already noticed, under the point of view of physical applicability, the above discussion can be useful in
describing the dynamics of propagating neutrinos which non-minimally couple with an approximately constant
external magnetic field {\boldmath$B$}.
In fact, the advent of the neutrino physics \cite{Zub98,Sch03,Alb03}
has stirred up an increasing number of theoretical works where
the quantum oscillation mechanisms \cite{Beu03,Vog04,Giu98,Giu02,Ber05A} have been deeply analyzed.
In particular, the correspondence between helicity and chirality has frequently caused some confusion in the literature
where the concept of helicity has been
erroneously used in the place of chirality for particles with $m = 0$.
In fact, some conceptual differences between the chiral and helicity operator has been analyzed \cite{Ahl06,Rol06}
in the context of the dual-helicity eigenspinors of the charge conjugation operator
(Elko)\footnote{In principle, a right-handed neutrino may be Elko, because it does not have a charge with respect
to any of the Standard Model gauge groups. When coupling to the left-handed sector with Yukawa-like
terms the coupling constant in such terms has positive mass dimension $1/2$.
Power counting suggests that quartic Elko terms may also be of importance.
These considerations may be of relevance for the understanding of neutrino
oscillations and neutrino mass generation, and to their relationship with dark matter and dark energy.}.

In the standard model of flavor-changing interactions, neutrinos with positive
chirality are decoupled from neutrino absorbing charged weak currents \cite{DeL98}.
Consequently, such positive chirality neutrinos become sterile with respect to weak interactions.
By reporting to the formalism with Dirac wave packets \cite{Zub80,Ber04} which leads to the
study of chiral oscillations \cite{DeL98}
(in vacuum), with the formal procedures presented in this manuscript, one could 
obtain the {\em immediate} description of chiral oscillations and spin flipping 
in terms of the Hamiltonian (\ref{08}) dynamics by recurring to the following
dynamic equation
\begin{eqnarray} 
\frac{d~}{dt}\langle \gamma^5 \rangle &=& 2\,i \langle\gamma_{\0}\,\gamma_{\5} \left[m - Re(\mu) \mbox{\boldmath$\Sigma$}\cdot \mbox{\boldmath$B$}\right]\rangle
\label{33}
\end{eqnarray} 
and
\begin{eqnarray} 
\frac{d~}{dt}\langle h \rangle &=& \frac{1}{2}\,Re(\mu)\langle \gamma_{\0} (\mbox{\boldmath$\Sigma$}\times \mbox{\boldmath$B$})\rangle \cdot \mbox{\boldmath$\hat{p}$}
\label{34},
\end{eqnarray} 
where the particle helicity is defined as 
the projection of the spin angular momentum onto the vector momentum direction
$h = \frac{1}{2}\mbox{\boldmath$\Sigma$}\cdot{\mbox{\boldmath$\hat{p}$}}$.
These results are obtained by means of a couple of very simple calculations,
nevertheless, we believe that, in spite of its simplicity, it is important in the large
context of quantum oscillation phenomena.
We still remark that, in the standard treatment of
vacuum neutrino oscillations, the use of scalar mass-eigenstate wave packets made up
exclusively of positive frequency plane-wave solutions is usually (implicitly)
assumed. Although the standard oscillation formula could give the correct
result when {\em properly} interpreted,
a satisfactory description
of fermionic (spin one-half) particles requires the use of the
Dirac equation as evolution equation for the mass-eigenstates.
Consequently, the spinorial form and
the interference between positive and negative frequency
components of the mass-eigenstate wave packets lead to the possibility
of chiral coupled with flavor oscillations \cite{Ber05A} which, eventually,
deserves a very careful investigation for cases of neutrinos which propagate through external interacting (magnetic) fields.

To summarize, we would have not been honest if,
by considering an arbitrary spatial configuration for the magnetic field,
we had ignored the complete analysis of the general case comprised by
Eqs.~(\ref{10}-\ref{12}). 
Meanwhile, such a peculiar observation leads to the formal connection between quantum oscillation phenomena and
a very different field. It concerns with the curious fact that the above complete (general) expressions
for propagating wave packets
do not satisfy the standard dispersion relations like $E^{\2} = m^{\2}+{\mbox{\boldmath$p$}}^{\2}$ excepting by the two
particular cases where $E_{\s}\bb{\mbox{\boldmath$p$}}^{\2} = m_{\s}^{\2}+ {\mbox{\boldmath$p$}}^{\2}$ for $\mbox{\boldmath$p$}\times\mbox{\boldmath$B$} = 0$
or $\epsilon_{\0}^{\2} = m^{\2}+ {\mbox{\boldmath$p$}}^{\2}$ for $\mbox{\boldmath$p$}\cdot\mbox{\boldmath$B$} = 0$.
By principle, it could represent an inconvenient obstacle which forbids the extension of these restrictive cases to
a general one. However, we believe that it can also represent a starting point in discussing 
dispersion relations which may be incorporated into frameworks encoding the breakdown (or the validity) of Lorentz invariance. 

\begin{acknowledgments}
The author thanks FAPESP (PD 04/13770-0) for Financial Support.
\end{acknowledgments}


\begin{thebibliography}{99}

\bibitem{Dir28}
P. A. M. Dirac, Proc. R. Soc. {\bf A117}, 610 (1928).
\bibitem{Esp99}
G. Esposito and P. Santorelli, J. Phys. {\bf A32}, 5643 (1999)
\bibitem{Alh05}
A. Alhaidari, hep-th/0503208.
\bibitem{Ein81}
E. Einchten and F. Feinberg, Phys. Rev. {\bf D23}, 2724 (1981)
\bibitem{Bak95}
M. Baker, J. S. Ball and F. Zachariasen, Phys. Rev. {\bf D51}, 1968 (1995)
\bibitem{Oli90}
J. C. D'Olivo, J. F. Nieves and P. B. Pal, Phys. Rev. Lett {\bf 64}, 1088 (1990).
\bibitem{Vol81}
M. B. Voloshin, M. I. Vysotskii and L. B. Okun, Zh. Eksp. Teor. Fiz. {\bf 91}, 754 (1986).
\bibitem{Ana98}
C.Athanassopoulos {\it et al.}, {\em ``LSND Collaboration''}, Phys. Rev. Lett {\bf 81}, 1774 (1998).
\bibitem{Agu01}
A. Aguilar {\it et al.}, {\em ``LSND Collaboration''}, Phys. Rev. {\bf D64}, 112007 (2001).
\bibitem{Ban03}
A. Bandyopadhyay {\it et al.}, Phys. Lett {\bf B583}, 134 (2004).
\bibitem{Zub80}
C. Itzykinson and J. B. Zuber, {\em Quantum Field Theory}, (Mc Graw-Hill Inc., New York, 1980).
\bibitem{Ber04}
A. E. Bernardini and S. De Leo, Eur. Phys. J.{\bf C37}, 471 (2004).
\bibitem{Kim93}
C. W. Kim and A. Pevsner, {\em Neutrinos in Physics and Astrophysics}, (Harwood Academic Publishers, Chur, 1993).
\bibitem{Gla61}
S. L. Glashow, Nucl. Phys. {\bf 20}, 579 (1961),\\
S. Weinberg, Phys. Rev. Lett. {\bf 19}, 1264 (1967),\\
A. Salam, {\em Elementary Particle Theory}, (N. Svartholm, Stocholm, 1968), p.367.
\bibitem{Ber05A}
A. E. Bernardini and S. De Leo, Phys. Rev. {\bf D71}, 076008-1 (2005)
\bibitem{Zub98}
K. Zuber, Phys. Rep. {\bf 305}, 295 (1998).
\bibitem{Sch03}
K. Scholberg, hep-ex/0308011
\bibitem{Alb03}
W. M. Alberico and S. M. Bilenky, Prog. Part. Nucl. {\bf 35}, 297 (2004).
\bibitem{Beu03}
M. Beuthe, Phys. Rep. {\bf 375}, 105 (2003).
\bibitem{Giu98}
C. Giunti and C. W. Kim, Phys. Rev. {\bf D58}, 017301 (1998).
\bibitem{Giu02}
C. Giunti, {\em JHEP} {\bf 0211}, 017 (2002).
\bibitem{Vog04}
R. D. McKeown and P. Vogel, Phys. Rep. {\bf 395}, 315 (2004).
\bibitem{Ahl06}
D. V. Ahluwalia-Khalilovaa and D. Grumiller, JCAP {\bf 0507}, 012 (2005).
\bibitem{Rol06}
R. da Rocha and W. A. Rodrigues Jr., Mod. Phys. Lett. {\bf A21}, 65 (2006).
\bibitem{DeL98}
S. De Leo and P. Rotelli, Int. J. Theor. Phys. {\bf 37}, 2193 (1998).

\end{thebibliography}
\end{document}